**[Title]** ORCID-linked labeled data for evaluating author name disambiguation at scale

**[Authors]** Jinseok Kim and Jason Owen-Smith

**[Author Information]**

Jinseok Kim (Corresponding Author)

Institute for Research on Innovation & Science, Survey Research Center, Institute for Social Research, University of Michigan
330 Packard Street, Ann Arbor, MI U.S.A. 48104-2910
734-763-4994|jinseokk@umich.edu

Jason Owen-Smith
Department of Sociology, Institute for Social Research, University of Michigan
330 Packard Street, Ann Arbor, MI U.S.A. 48104-2910
734-936-0463|jdos@umich.edu

Abstract

How can we evaluate the performance of a disambiguation method implemented on big bibliographic data? This study suggests that the open researcher profile system, ORCID, can be used as an authority source to label name instances at scale. This study demonstrates the potential by evaluating the disambiguation performances of Author-ity2009 (which algorithmically disambiguates author names in MEDLINE) using 3 million name instances that are automatically labeled through linkage to 5 million ORCID researcher profiles. Results show that although ORCID-linked labeled data do not effectively represent the population of name instances in Author-ity2009, they do effectively capture the 'high precision over high recall' performances of Author-ity2009. In addition, ORCID-linked labeled data can provide nuanced details about the Author-ity2009's performance when name instances are evaluated within and across ethnicity categories. As ORCID continues to be expanded to include more researchers, labeled data via ORCID-linkage can be improved in representing the population of a whole disambiguated data and updated on a regular basis. This can benefit author name disambiguation researchers and practitioners who need large-scale labeled data but lack resources for manual labeling or access to other authority sources for linkage-based labeling. The ORCID-linked labeled data for Author-tiy2009 are publicly available for validation and reuse.

Keywords: author name disambiguation; labeled data; ORCID; record linkage; Author-ity2009

## Introduction and Background

Many author name disambiguation studies have evaluated the performances of their proposed methods on truth data labeled by human coders (e.g., Han, Giles, Zha, Li, & Tsioutsiouliklis, 2004; Qian, Zheng, Sakai, Ye, & Liu, 2015; Santana, Gonçalves, Laender, & Ferreira, 2017; X. Wang, Tang, Cheng, & Yu, 2011). Generating manually labeled data is, however, a daunting challenge. Given the same queues of name instances, for example, human coders can disagree up to 25% of cases (e.g., Liu et al., 2014; Smalheiser & Torvik, 2009). In addition, labeling decisions agreed upon by human coders can be wrong (J. Kim, 2018; Shin, Kim, Choi, & Kim, 2014). Mostly importantly, manual labeling is not scalable. Labeling a few thousand name instances can take several months (Kang, Kim, Lee, Jung, & You, 2011) or require multiple verification steps (Song, Kim, & Kim, 2015), which is labor-intensive and time-consuming. So, manual labeling is often unsuitable for evaluating a disambiguation task handling a large number of name instances.

In an effort to avoid the limitations of manual labelling, several studies have created labeled data without human coders. For example, Torvik and Smalheiser (2009) labeled name instances sharing the same email addresses as representing the same author. To decide whether name instances refer to the same author or not, other studies used different features of publication data such as shared coauthors (e.g., Cota, Ferreira, Nascimento, Gonçalves, & Laender, 2010) or self-citation (e.g., Levin, Krawczyk, Bethard, & Jurafsky, 2012). These labeling methods produce labels at large scale (up to millions of labeled instances) but their labeling results have rarely been verified for accuracy.[1] As they are designed to produce positive (i.e., label match) sets of name instance pairs, they often require negative (i.e., label nonmatch) sets generated by heuristic rules (e.g., name instances with different name string and no shared coauthors are assumed to refer to different authors). To correct this problem, an iterative clustering method that triangulates multiple matching features such as coauthors, email addresses, and self-citation has been proposed. But its effectiveness can be constrained if those discriminating features are poorly recorded for a given set of name instances (J. Kim, Kim, & Owen-Smith, 2019).

Another group of studies has relied on third-party data sources that control the accuracy of researcher information. For example, Kawashima and Tomizawa (2015) evaluated the disambiguation performance of SCOPUS on a list of 75,405 Japanese author names in 573,338 papers. For this, they used the Database of Grants-in-Aid for Scientific Research (KAKEN) that maintains a unique ID number of a funded researcher in Japan with a list of her/his verified publications. An author name instance in a SCOPUS-indexed paper was compared to each KAKEN researcher profile by comparing name strings, publication records, and affiliations. If a match was found, the KAKEN researcher ID was assigned to the author name instance. Such a record linking technique has been used in other studies to label name instances of Italian researchers (D'Angelo, Giuffrida, & Abramo, 2011) and Dutch researchers (Reijnhoudt, Costas, Noyons, Borner, & Scharnhorst, 2014) using each nation's administrative scholarly databases. Other sources for labeling include NIH-funded researcher profiles[2] (e.g., K. Kim, Sefid, Weinberg, & Giles, 2018; Lerchenmueller & Sorenson, 2016; Liu et al., 2014; Torvik & Smalheiser, 2009) and Highly Cited Researchers data[3] (e.g., Liu et al., 2014; Torvik & Smalheiser, 2009). While these record linkage procedures produce large-scale, accurate labeling results, it also provides biased results (Lerchenmueller

---

[1] An exception is Levin et al. (2012) in which name instances that match on email addresses are verified by authors through email correspondence.
[2] https://exporter.nih.gov/
[3] https://hcr.clarivate.com/

& Sorenson, 2016). For example, name instances of researchers who are not active in a targeted nation or discipline, not funded by NIH, or not highly cited cannot be labeled.

To address the problems, a few studies have recently begun to use the Open Researcher & Contributor ID (ORCID)[4] data as an authority source to label name instances for disambiguation evaluation (e.g., J. Kim, 2018; J. Kim, 2019b; J. Kim et al., 2019). Similarly, several studies have discussed the potential of using ORCID for authority control within and across digital libraries (e.g., Francis, 2013; Mallery, 2016; Thomas, Chen, & Clement, 2015). ORCID is an open platform of more than 5 million researcher profiles curated by individual researchers for education history, authorship, and employment information (Haak, Fenner, Paglione, Pentz, & Ratner, 2012). Like other authority sources mentioned above, linking ORCID to bibliographic data can produce large-scale labeled data of up to one million instances (J. Kim, 2019b). Unlike other sources, however, author profiles in ORCID are not limited to specific disciplines, geographic regions, organizations, or high-visibility scholars. This implies that ORCID has a potential to label names of researchers from diverse backgrounds and thereby overcome the limited coverage of other authority sources.

But the potential benefits of ORCID for this task have be insufficiently analyzed. A few questions can be asked to characterize labeling results through ORCID-linkage:

(1) How well do ORCID-linked labeled data represent the population of name instances in a large-scale bibliographic dataset?

(2) How do ORCID-linked labeled data compare to other labeled data generated by different methods?

(3) What are the benefits and cautions that must be considered before ORCID is used as a labeling source for evaluating author name disambiguation?

The answers to these questions can help disambiguation researchers to make informed choices of labeled data and to create evaluation and ground-truth datasets at scale. Several studies have attempted to answer similar questions by discussing how ORCID profiles represent the author population in Web of Science (Youtie, Carley, Porter, & Shapira, 2017), what issues need to be addressed before ORCID can be used as a gold standard for author disambiguation (Albusac, de Campos, Fernández-Luna, & Huete, 2018; Eichenlaub & Morgan, 2017), and how record-linkage-based labeling may or may not work in author disambiguation under certain conditions (Anderson A Ferreira, Gonçalves, & Laender, 2020; Reijnhoudt et al., 2014). This study contributes to that growing literature by demonstrating the use of ORCID-linked labeling against another large-scale disambiguated dataset constructed using different linkage-based labeling methods. Specifically, this study labels name instances in MEDLINE by linking them with ORCID researcher profiles. Then, the performances of Author-ity2009 which disambiguates MEDLINE author names, is evaluated using the labeled data. For comparison, two labeled datasets are created using two widely-used sources - NIH-funded researcher information and self-citation information. The three labeled datasets are compared for their representativeness of Author-ity2009 as well as to evaluate results of the Author-ity2009's disambiguation performances. After that, a discussion follows about the implications and challenges of using ORCID for labeling. In the following section, labeling procedures via record-linkage for Author-ity2009 are described in detail.

## Methodology

### Author-ity2009: Evaluation Target

---

[4] https://orcid.org/

This study shows the potential of ORCID-linkage-based labeling for evaluating author name disambiguation by assessing the disambiguation performance of Author-ity2009 (Torvik & Smalheiser, 2009; Torvik, Weeber, Swanson, & Smalheiser, 2005). Author-ity2009 is a bibliographic database that contains disambiguated author names in MEDLINE[5], the world's largest digital library of biomedical research, maintained by the U.S. National Library of Medicine (NLM). In Author-ity2009, author names are disambiguated in two steps. First, name pairs are compared for similarity over various features such as middle name initial, coauthor name, affiliation, and Medical Subject Headings. Next, the instance pairs are grouped into clusters by a maximum-likelihood-based, hierarchical agglomerative clustering algorithm using the pairwise similarity calculated in the first step.

Author-ity2009 is chosen as an evaluation target for three reasons. First, Author-ity2009 conducts author name disambiguation on a digital library scale: 61.7M name instances in 18.6M papers published between 1966~2009 as indexed in MEDLINE. Evaluating disambiguation results for such a large bibliographic corpus can be a daunting challenge. So, Author-ity2009 can be a good use case to illustrate how ORCID-linkage can contribute to the performance and evaluation of an important, large-scale disambiguation task. Second, the performance of Author-ity2009 have been evaluated on different types of labeled data in several studies (e.g., J. Kim, 2019b; Lerchenmueller & Sorenson, 2016; Liu et al., 2014; Torvik & Smalheiser, 2009), as summarized in Table 1. This provides a context for comparing ORCID with other labeling sources to better understand its strengths and weaknesses. Third, Author-ity2009 is publicly available for research, enabling scholars to replicate and validate this study.

*Table 1: Summary of Labeled Data in Selected Studies Evaluating Author-ity2009*

| Reference | Labeling Method | Labeled Data |
|---|---|---|
| **Torvik and Smalheiser (2009)** | Manual | Papers of 62 randomly selected author names |
| | Automatic | 323,274 self-citation pairs |
| | Linkage | 20,085 researcher profiles in Community of Science<br>2,313 Highly Cited Researcher profiles in Web of Science<br>83,992 NIH-funded PI information |
| **Liu et al. (2014)** | Manual | 300 randomly selected pairs of author name |
| | Automatic | 4.7 million self-citation pairs<br>23 million grant-citation pairs |
| | Linkage | 40 Highly Cited Researcher profiles in Web of Science<br>47 NIH-funded PI information |
| **Lerchenmueller and Sorenson (2016)** | Linkage | 36,987 NIH-funded PI information |
| **K. Kim et al. (2018)** | Linkage | 54,260 NIH-funded PI information |
| **J. Kim (2019b)** | Linkage | 130,712 ORCID researcher profiles |

Files containing disambiguated names in Author-ity2009 (Torvik & Smalheiser, 2018) are downloaded from Illinois Data Bank[6]. A unique author in the Author-ity2009 file is represented by an author ID with a list of name instances of the author. A name instance is represented by an instance ID which is a numeric combination (e.g., 1234567_2) of (1) PMID (7~8 digit numbers) of a paper in which the instance appears and (2) the instance's byline position (1, 2, 3 … *N*) in the paper. The downloaded Author-ity2009 contains a total of 61.7 M name instances in 18.6 M papers.

---
[5] https://www.nlm.nih.gov/bsd/medline.html
[6] https://databank.illinois.edu/datasets/IDB-4222651

MED-ORC: Linking MEDLINE with ORCID

To evaluate the disambiguation performances of Author-ity2009, author name instances disambiguated by Author-ity2009 need to be labeled. This study attempts to link ORCID ids to 40M author name instances that appear in about 9M papers published between 1991 and 2009 in Author-ity2009. Author-ity2009 disambiguates author name instances in MEDLINE but does not provide their raw name strings. So, this study proceeds from the whole MEDLINE corpus (2016 baseline version) retrieved from the National Library of Medicine repository[7]. We select MEDLINE records for papers published between 1991 and 2009 (MEDLINE2009) to align with the publication year range of Author-ity2009. Next, name instances in MEDLINE2009 are compared to the author profiles in ORICD. For this MEDLINE2009-ORCID linkage, a 2018 ORCID release version is used[8]. To find author name instances recorded in both MEDLINE2009 and ORCID, paper titles with five or more words in MEDLINE2009 are encoded into ASCII format, deprived of non-alphabetical characters, and lowercased. Any duplicate titles after the pre-processing are removed. Then, each title (which is associated with a unique PMID) is compared to the publication lists in ORCID researcher profiles. If a match is found between bibliographic records in MEDLINE2009 and ORCID, author name strings that appear in the matched MEDLINE2009 paper are compared with the name string of the ORCID researcher whose list of publications contains the matched title. If two name strings in MEDLINE2009 and ORCID are matched on the full surname plus the first forename initial, they are assumed to refer to the same author and the ORCID ID of the matched researcher profile is assigned to the name instance in MEDLINE2009. As shown in Figure 1, this matching process produces a labeled dataset, MED-ORC, in which an author name instance in a MEDLINE paper is associated with an ORCID ID.

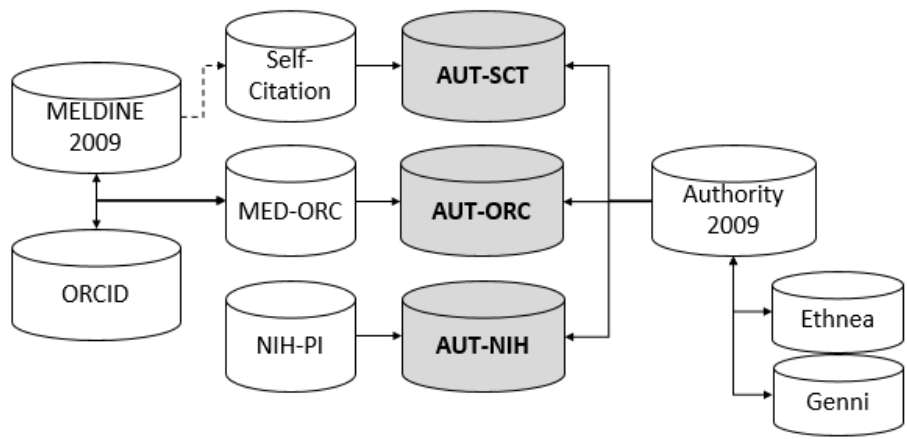

*Figure 1: An Overview of Data Linkage and Outcome Data for Analysis*

AUT-ORC: Linking Author-ity2009 with MED-ORC

Author-ity2009 is linked to MED-ORC to create a subset of Author-ity2009 (AUT-ORC) in which an author name in a MEDLINE paper is associated with both (1) an author label assigned through MEDLINE2009-ORCID linkage and (2) an Author-ity2009 ID assigned through the disambiguation conducted by Torvik and Smalheiser (2009). The resulting data, AUT-ORC, contain 3,076,501 author name instances, which we used to assess the disambiguation performance of Author-ity2009. Table 2 shows an example of a data instance in AUT-ORC in which an author name in a MEDLINE paper is associated with a PMID, byline position, name string, Author-ity2009 ID, and ORCID ID.

---

[7] ftp://ftp.ncbi.nlm.nih.gov/pubmed/baseline. This study used the 2016 baseline.
[8] https://figshare.com/articles/ORCID_Public_Data_File_2018/7234028

*Table 2: An Illustration of Data Instance in AUT-ORC Linking Author-ity2009 and ORCID*

| PMID | Author Position | Name String | Author-ity2009 ID | ORCID ID | Ethnicity | Gender | Publication Year |
|---|---|---|---|---|---|---|---|
| 1701372 | 1 | Hertzog, P J | 124851_1 | 0000-0002-XXXX-XXXX | English | Male | 1991 |

To better understand the composition of ORCID-linked Author-ity2009 name instances, this study uses name ethnicity and gender information as illustrated in Table 2. A name instance in Author-ity2009 is assigned a name ethnicity tag by an ethnicity classification system, *Ethnea*, developed by Torvik and Agarwal (2016)[9]. *Ethnea* assigns one of 26 name ethnicity tags to an author name instance in Author-ity2009 based on the name's association with national-level geo-locations[10]. For example, "Wei Wang" is tagged as 'Chinese' because it is most frequently associated with China-based organizations. Meanwhile, the gender of a name instance is obtained from *Genni*, a gender prediction tool developed also by Dr. Torvik's team at the University of Illinois at Urbana-Champaign[11]. *Genni* assigns one of three gender categories – Female, Male, and Unknown – to a name instance based on its first name's association with frequent gender signifiers (e.g., my aunt Taylor) in combination of its surname's ethnicity (e.g., 'Andrea' can be male or female depending on regions where it is used[12]).

AUT-NIH: Linking Author-ity2009 with NIH PI Data

This study also creates a benchmark labeled dataset by linking Author-ty2009 with Principal Investigator (PI) information recorded in the National Institutes of Health (NIH) funded research data (ExPORTER). This NIH-linkage has been used in several studies to evaluate author name disambiguation for MEDLINE because ExPORTER provides the PMIDs of research papers in MEDLINE that result from NIH funds (e.g., K. Kim, Sefid, & Giles, 2017; K. Kim et al., 2018; Liu et al., 2014; Torvik & Smalheiser, 2009). After an Author-ity2009 paper's PMID is found to be associated with a specific NIH grant, the author names in the paper are compared to the names of the PI who received the funding. If a PI's name is found to match an author name, her/his unique NIH PI ID is assigned to the author name as a label. This study reuses the list of NIH PI IDs linked to the Author-ity2009 in Lerchenmueller and Sorenson (2016)[13]. To make this NIH-linked labeled data (AUT-NIH) comparable to AUT-ORC, each name instance in AUT-NIH is assigned an ethnicity and a gender using *Ethnea* and *Genni* each.

AUT-SCT: Linking Author-ity2009 with Self-Citation Information

---

[9] https://databank.illinois.edu/datasets/IDB-9087546
[10] 26 ethnicities include: African, Arab, Baltic, Caribbean, Chinese, Dutch, English, French, German, Greek, Hispanic, Hungarian, Indian, Indonesian, Israeli, Italian, Japanese, Korean, Mongolian, Nordic, Polynesian, Romanian, Slav, Thai, Turkish, and Vietnamese. Some name instances are assigned compound ethnicities (e.g., "Jane Kim" → Korean-English) if the surname and forename of an author name are associated frequently with different ethnicities.
[11] Genni + Ethnea for the Author-ity 2009 dataset. (2018). Retrieved from: https://doi.org/10.13012/B2IDB-9087546_V1
[12] https://en.wikipedia.org/wiki/Andrea
[13] https://dx.doi.org/10.6084/m9.figshare.3407461.v1. Instead of 355K instances in the original linked data, this study filters 313K instances recorded in papers published between 1991 and 2009.

Another benchmark labeled dataset is a list of name instance pairs that represent self-citation relations. This self-citation information has been used in several studies to develop and test automatic labeling methods (e.g., J. Kim, 2018; J. Kim et al., 2019; Liu et al., 2014; Schulz, Mazloumian, Petersen, Penner, & Helbing, 2014; Torvik & Smalheiser, 2009). This labeling method is based on the assumption that if a paper cites another and they have the same author names, those names refer to the same author. To generate a list of citing references for a paper, reference lists of papers in MELDINE are connected to their cited papers via matching PMIDs. Then, author names in a cited paper are compared to those in citing papers. Following the common practice using this labeling method, if two name instances in cited and citing papers each match on the full surname and the first forename initial, we treat them as instances of the same author. More than 6.2M self-citation pairs are detected in Author-ity2009. To be comparable to AUT-ORC and AUT-NIH, each name instance in a self-citation pair is assigned an ethnicity and a gender, too. Table 3 characterizes the sources of record linkage and labeling methods of the three labeled datasets – AUT-ORC, AUT-NIH, and AUT-SCT – and presents the numbers of labeled instances and unique authors in each dataset. Note that the number of unique authors is unavailable for AUT-SCT because only name instances that have self-citation relationships can be labelled. It is thus impossible to know from this dataset alone whether name instances without self-citation refer to the same author.[14]

*Table 3: Summary of Three Labeled Data Used for Comparison*

| Linked Data | Linkage Sources | Labeling Method | Number of Labeled Instances | Number of Unique Authors |
|---|---|---|---|---|
| **AUT-ORC** | Author-ity2009 & ORCID | Paper title match + Author Name Match | 3,076,501 | 245,755 |
| **AUT-NIH** | Author-ity2009 & NIH ExPORTER | Paper PMID match + Author Name Match | 312,951 | 34,206 |
| **AUT-SCT** | Author-ity2009 & Reference lists in MEDLINE | Paper PMID match + Author Name Match | 6,214,199 (Name pairs) | - |

Performance Evaluation

*Clustering Measure*: To assess the performances of Author-ity2009 on three labeled datasets, author name instances referring to the same author are grouped into a cluster. Specifically, a truth cluster is the collection of author name instances that share the same ORCID ID (AUT-ORC) or the same NIH PI ID (AUT-NIH). Meanwhile, a predicted cluster is the collection of author name instances that share the same Author-ity2009 ID (AUT-ORC and AUT-NIH). Then, the predicted cluster is compared to the truth cluster to quantify how well it contains only and all instances that belong to the truth cluster. This study uses B-Cubed ($B^3$), one of most frequently used clustering metrics in author name disambiguation (J. Kim, 2019a). This measure is comprised of three metrics: $B^3$ Recall, $B^3$ Precision, and $B^3$ F1, which are defined as follows:

$$B^3 Recall = \frac{1}{N} \sum_{t \in T} \frac{|P(t) \cap T(t)|}{|T(t)|} \quad (7)$$

---

[14] For example, let's assume that two pairs, A-B and C-D, are in self-citation relation. If the pair of B-C is in self-citation relation, then A, B, C, and D can be grouped into a cluster via transitivity, as illustrated in Schulz et al. (2014). But such information is not always available for all instance pairs in AUT-SCT.

$$B^3 \, Precision = \frac{1}{N} \sum_{t \in T} \frac{|P(t) \cap T(t)|}{|P(t)|} \quad (8)$$

$$B^3 \, F1 = \frac{2 \times R \times P}{R + P} \quad (9)$$

Here, *t* represents an author name instance in truth clusters *T*. *N* refers to the number of all author name instances in truth clusters (*T*). $T(t)$ is a truth cluster containing an author name instance *t*, while $P(t)$ a predicted cluster holding the name instance *t*.

*Classification Measure*: Author name instances in AUT-SCT are recorded as self-citation. We evaluate the disambiguation performance of Author-ity2009 by comparing two Author-ity2009 IDs associated with each of the two paired name instances. If they have the same IDs, Author-ity2009 succeeds in correctly classifying the pair as a matched set, while if IDs are not matched, it fails (→ binary classification). So, the performance of Author-ity2009 is quantified by calculating the ratio of truth pairs in self-citation that have the same Author-ity2009 IDs (≈ recall).

*Baselines*: Following previous studies (Backes, 2018; J. Kim, 2018; J. Kim & Kim, 2020; Louppe, Al-Natsheh, Susik, & Maguire, 2016), this study uses two heuristics as baseline methods for comparing how well Author-ity2009 performs in disambiguation. The first heuristic decides author name instances matched on the full surname and all forename initials to represent the same author (AINI hereafter). This method has been used by many bibliometric scholars for decades and as one of the standard name formats in major digital libraries (Garfield, 1969; Milojević, 2013; Strotmann & Zhao, 2012). Meanwhile, the second heuristic decides name instances to represent the same author if they share the full surname and the first forename initial (FINI hereafater). This method has also been used as both a disambiguation heuristic and an author name query format in digital libraries. In addition, most disambiguation studies use this method to group name instances that are disambiguated together (blocking). These two heuristics provide bottom-line performances to evaluate Author-ity2009.

Results

Representativeness

*Distribution of Publication Years*: As shown in Table 3, the three labeled datasets – AUT-ORC, AUT-NIH, and AUT-SCT –contain different numbers of labeled name instances (pairs). How do those instances differ and which is most representative of the overall Author-ity2009 dataset? To characterize the composition of labeled data, publication years of papers in which a labeled instance appears are counted. Figure 2 compares the publication year distributions of three labeled datasets. Note that for AUT-SCT, years associated with each self-citing name instance pair are counted.

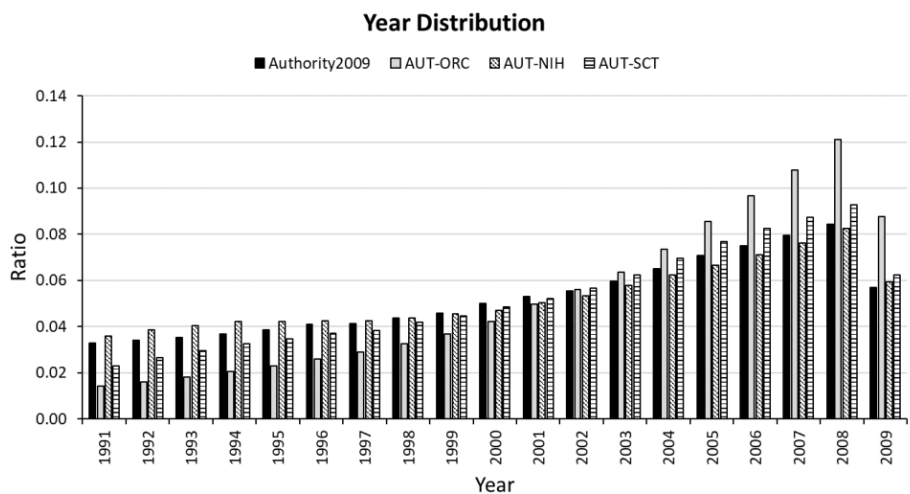

*Figure 2: Distribution of Publication Year Associated with Labeled Instances in Three Labeled Data*

Figure 2 shows that in Author-ity2009 (black bar), the number of author name instances consistently increases over time[15], which aligns with secular growth trends in the overall number of publications (more papers generally result in more author name instances) and with increasing team sizes in science (Bornmann & Mutz, 2015; Wuchty, Jones, & Uzzi, 2007). Compared against Author-ity2009, the year distribution of AUT-ORC (gray bar) shows that instances in papers published before 2002 constitute a smaller proportion of all instances in AUT-ORC than in Author-ity2009. For example, AUT-ORC instances associated with publications in 1991 are 1.43 percent of all instances in AUT-ORC, while the Autor-ity2009 instances for the same year make up 3.30 percent of all instances in Autor-ity2009. This trend is reversed after 2002: in AUT-ORC, instances appearing in a specific year make up more proportion of all instances than those in the Author-ity2009. For example, the percentage of AUT-ORC instances in 2008 is 12.11(%), while the percentage of Author-ity2009 instances in the same year is 8.44(%). This indicates that in AUT-ORC, name instances that appear in more recently published papers are over-represented compared to the publication year distribution of Author-ity2009 name instances. This pattern is consistent with the recent growth of ORCID and with the greater likelihood that early and mid-career researchers will have and actively maintain ORCID profiles than older and more established researchers (Youtie et al., 2017).

The year distribution of AUT-NIH (diagonal-line bar) shows a very similar pattern to that of Author-ity2009. Interestingly, AUT-SCT (horizontal-line bar) has a similar pattern to that of AUT-ORC: roughly before and after 2002, the ratios of instances with a specific year are lower and higher than those of instances in Author-ity2009, respectively. Assuming that the tendency of self-citation among scholars does not change much over time, this pattern may arise from the combination of two trends: (1) scholars in the sciences tend to cite more recent papers, usually focusing on those published within 5 years (Tahamtan, Safipour Afshar, & Ahamdzadeh, 2016; J. Wang, 2013); and (2) more papers have been published in recent years (Bornmann & Mutz, 2015) as the overall growth of the scientific literature has accelerated. Another possible explanation might be that self-citation itself has become more common over time. But validating that possibility is beyond the scope of the present study. Overall Figure 2 indicates that AUT-NIH most closely matches Author-ity2009 in terms of the publication year distribution of name instances. The other two labeled datasets over-represent recent years heavily (AUT-ORC) and slightly (AUT-SCT) relative to Author-ity2009..

---

[15] An exception is the year of 2009 when the number of publication decreased because of the incomplete coverage of the Author-ity2009. This incompleteness seems to be caused by publishers who submitted publication records for 2009 to MEDLINE later than 2009 after their internal record processing and quality control.

*Gender Distribution*: To provide another indicator of how well each labelled dataset represents Author-ity2009, we turn to comparisons of the gender composition of author name instances. Figure 3 shows that the majority of name instances in all datasets are male (black bar; 57%) while female instances (22.32%) and NULL (i.e., gender unidentifiable) instances (20.28%) make up the rest with similar percentages. Such an imbalanced gender distribution is broadly characteristic of scientific authorship in general (Larivière, Ni, Gingras, Cronin, & Sugimoto, 2013) as of biomedical science (Jagsi et al., 2006). The gender imbalance is also observed in AUT-ORC (gray bar) in which male names constitute 67.46 percent of all name instances while the percentage of female instances (22.70%) is quite similar to that in Autor-ity2009. The higher ratio of male instances in AUT-ORC than in Author-ity2009 seems to be a trade-off with the reduced ratio of Null name instances. The same pattern is observed in AUT-NIH and AUT-SCT in which the dominance of male names are more prevalent (i.e., 73.95% and 65.44% each) than in Autor-ity2009 and AUT-ORC but with lower ratios of Null names and similar ratios of female names. These observations indicate that despite the minute differences in gender ratios, three labeled data shared similar patterns of gender distribution.

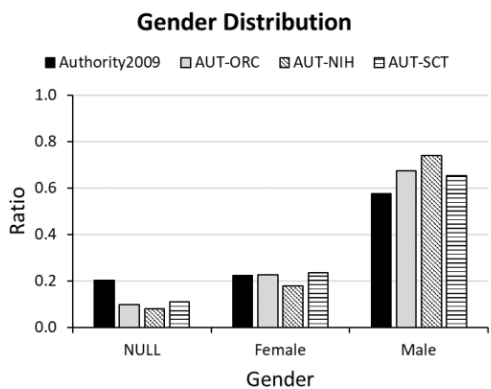

*Figure 3: Distribution of Gender Associated with Labeled Instances in Three Labeled Data*

*Ethnicity Distribution*: Several studies have investigated how name ethnicities are distributed as a means to characterize labeled data (e.g., J. Kim et al., 2019; Lerchenmueller & Sorenson, 2016). According to Figure 4, the largest ethnic group in Author-ity2009 (black bar) is English (24.36%), followed by Japanese (10.15%), German (8.39%), Chinese (7.60%), and Hispanic (6.78%). The largest ethnic group in AUT-ORC (gray bar) is also English (24.66%) whose ratio is very close to that in Author-ity2009. Unlike Author-ity2009, however, the second largest group in AUT-ORC is Hispanic (14.16%), followed by Italian (12.05%). This disparity can be attributed to the fact that researcher profiles in ORCID are disproportionally associated with European countries, especially Italy and Spain (Youtie et al., 2017). In contrast, Asian names (e.g., Japanese, Chinese, and Korean) are under-represented in AUT-ORC as they are in ORCID (J. Kim et al., 2019; Youtie et al., 2017).

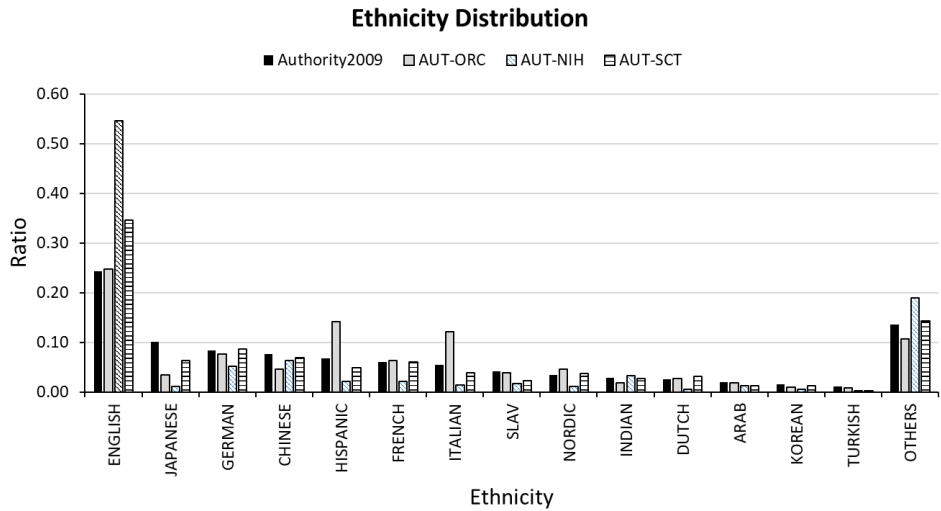

*Figure 4: Distribution of Ethnicity Associated with Labeled Instances in Three Labeled Data*

English name instances constitute the majority in AUT-NIH (diagonal-line bar, 54.53%), while other ethnicities are heavily underrepresented compared to their ratios in Author-ity2009. This might be because AUT-NIH is created based on information of PIs who have ever received funds from NIH in the U.S. Non-US investigators are generally ineligible to apply for NIH funds, so it makes sense that the name instance distribution in this dataset would skew toward English names. This English-skewed distribution is also confirmed in Lerchenmueller and Sorenson (2016) who found that 84% of all ethnicity-identified instances in the *whole* Author-ity2009 linked to NIH ExPORTER are 'Caucasian' (including many European names as well as English). Meanwhile, many instances in self-citation relation are also English (horizontal-line bar; 34.65%) but the ratio differences of other ethnicities against Author-ity2009 are smaller compared to those in AUT-ORC and AUT-NIH. As such, three labeled data are common in that English name instances are prevalent but none of them represents well the ethnicity distribution in Author-ity2009 because some ethnicities are over-represented while others under-represented.

*Block Size Distribution*: Another way to discover how three labeled datasets represent Author-ity2009 is to compare the distributions of block sizes in each dataset. A common practice in author name disambiguation research is to collect author name instances into a block if they match on the full surname and first forename initial. Comparisons that support disambiguation are then performed within blocks (K. Kim et al., 2018). Many studies have used the block size distribution to characterize labeled data (e.g., J. Kim et al., 2019; Levin et al., 2012; Müller, Reitz, & Roy, 2017; Torvik & Smalheiser, 2009). Block sizes can become huge because labeled data contain a few hundreds of thousands (AUT-NIH) or millions (AUT-ORC) of name instances. So, block size distributions are plotted using a cumulative density function on log-log axes. Figure 5 visualizes the block size distributions in AUT-ORC and AUT-NIH. Note that AUT-SCT cannot produce a block size distribution because self-citation pairs only contain match information at the pair level and, thus, the matching status of name instance pairs that are not in self-citation relation but that may nevertheless fall within a block defined by surname and first initial are still unknown.

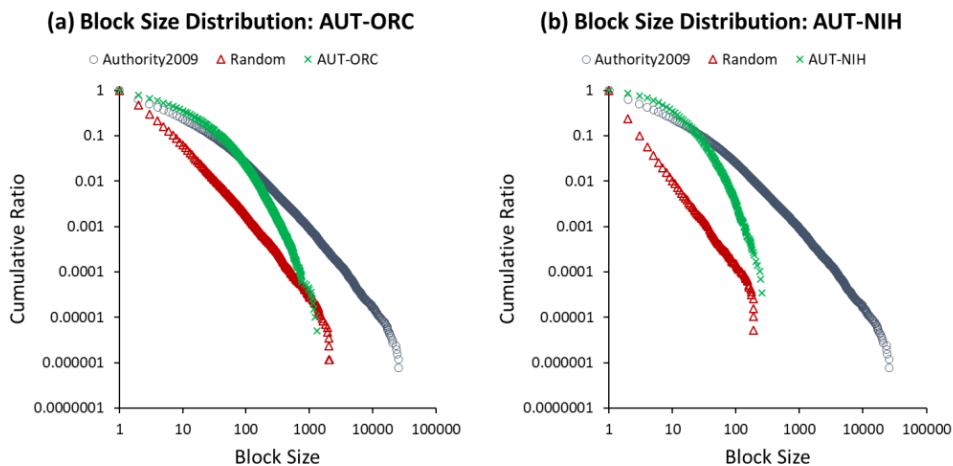

*Figure 5: Comparison of Block Size Distributions in Labeled Data against Author-ity2009*

In Figure 5, the *x*-axis shows block sizes ranging from 1 (i.e., a single instance block) to 25,917 in Author-ity2009 (Figure 1a). The *y*-axis represents the ratio of blocks with a specific size or larger (cumulative) over all blocks is shown. For example, in Author-ity2009, blocks with 2 or more name instances (blue circles) constitute 63.47% of all blocks, which in reverse means that 36.53% of blocks contain only one instance. In Figure 5a, the block size distributions of Author-ity2009 (blue circles) and AUT-ORC (green x-markers) are compared. Both distributions are highly skewed: most blocks are small while a few are huge. In Author-ity2009, for example, blocks with 12 or fewer instances make up 80% of all blocks, while in AUT-ORC, blocks with 20 or fewer do so. But they begin to differ as the block size increases over those 80% thresholds. The curvature of AUT-ORC turns downward more than that of Author-ity2009. This means that in AUT-ORC, large blocks make up a smaller proportion of AUT-ORC than they do inAuthor-ity2009. To see if this difference naturally occurs due to different data sizes (AUT-ORC ≈ 3M vs Author-ity2009 ≈ 40M), we randomly select a set of Athor-ity2009 name instances of the same general size as AUT-ORC (i.e., 3M). Their block size distribution is depicted on the figure in red. As shown in Figure 5a, the random data's block size distribution (red triangles) has a different shape from AUT-ORC's, while it has a similar curvature as that of Author-ity2009. This implies that the block size distribution in AUT-ORC is biased toward small sizes when compared to its population data, Author-ity2009. The same pattern is also observed for AUT-NIH in Figure 5b. Both these distances may be due to the relatively larger European focus of ORCID and the US focus of NIH data. The largest name blocks in Author-ity2009 tend to be created by highly ambiguous Asian name instances. If the distributions of both labeled datasets were representative of Author-ity2009, we would expect to see their distributions track closely with those of the random subset of Author-ity2009 name instances.

Clustering Performance: AUT-ORC and AUT-NIH

How can we describe the disambiguation performance of Author-ity2009 evaluated on each labeled data? Figure 6 reports B-cubed ($B^3$) recall, precision, and F1 scores calculated on AUT-ORC and AUT-NIH in comparison with two baseline performances: AINI – all forename initials based disambiguation and FINI – first forename initial based disambiguation). According to Figure 6a, Author-ity2009 performs better than AINI but worse than FINI in finding all name instances associated with distinct authors (Recall). To be specific, the high recall by FINI is expected because only name instances within the same block are disambiguated and all name instances of the same author belong to the same block. As a block consists of name instances sharing the same full surname and first forename initial, matching author identities of instances based on their full surname and first forename initial is supposed to find all name instances of the same author (FINI ≈ blocking). But as seen in Figure 6a, even FINI fails to obtain a perfect recall score: The gray bar's height stops below 1. This means that in AUT-ORC, some instances have different 'full surname + first forename initial' formats while they actually refer to the same author (because they share the same ORCID ids). Among 245,755 unique authors in the AUT-ORC, 12,646 (5.15%) authors (273,782 instances) have at least one name string that has a different 'full surname + first forename initial' format with others. A semi-manual inspection reveals that they are synonyms with three different types[16]. First, about 77% of the 12,646 authors have their surnames recorded in different strings. For example, an author whose name is 'Wagner Luiz do Prado' has four name variants – 'Prado, Wagner L.'; 'do Prado, Wagner Luiz'; 'do Prado, W. L.'; and 'Prado, Wagner Luiz do' – in MEDLINE. While the author's name instances share the same first forename initial ('W'), his surnames are recorded in two different strings ('Prado' and 'do Prado'). This type also occurs when surnames of an author are recorded in different strings due to, for example, inconsistent encodings of special characters (e.g., López → Lpez). The next most frequent type (1,892 authors; 15%) is the case where first forename initials are different while surnames are the same. For example, an author whose name is 'Patricia Miang Lon Ng' is recoded in three different strings – 'Ng, Patricia Miang Lon,' 'Ng, Miang Lon Patricia,' and 'Ng, Patricia M. L.' When simplified into the full surname + first forename initial format, the author is represented by two different names – 'Ng, P,' and 'Ng, M.' The third type (973 authors; 8%) occurs when the order of surname and forenames is flipped (e.g., Wei, Wang → 'Wei, W.'; Wang, Wei → 'Wang, W.').

Second, the worse recall by AINI can be explained by the fact that matching name instances based on all forename initials (= AINI) cannot detect instances that have different forename initials but refer to the same author. For example, two instances of an author, 'Brown, C' and 'Brown, C. C.' are decided to refer to different authors by AINI. Meanwhile, Author-ity2009 decides whether name instances refer to the same author or not by calculating their similarity over several features such as name string, coauthor names, affiliation information, title words, etc. Such a sophisticated method can find more synonymous instances that belong to the same author than AINI but could not perform at par with the heuristic disambiguation of FINI.

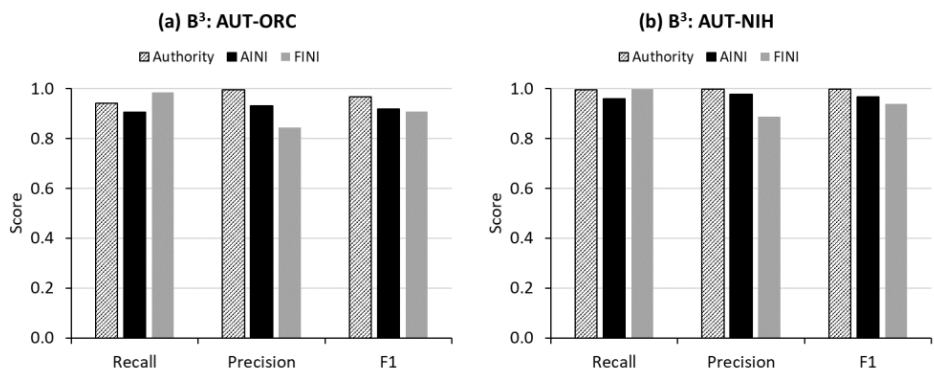

Figure 6: Evaluation of Disambiguation Performances of Author-ity2009 Evaluated on Two Labeled Data

---

[16] Among 12,646 authors (by unique ORCID ids) who have two or more 'full surname + first forename initial' formats, we randomly selected 100 authors and examine *manually* their 'full surname + first forename initial' formats to categorize them into three types: (1) different surname + same forename initial, (2) same surname + different forename initial, and (3) flipped name order. Then, using a script language (Perl), one of the three types were assigned *automatically* to the remaining 12,546 authors.

Author-ity2009 excels relative to the two baseline methods in precision (Figure 6a). Its precision score is almost perfect (0.99), whereas AINI records 0.93 and FINI 0.84. The high precision by Author-ity2009 means that in addition to correctly identifying name instances associated with the same authors (precision), it can also distinguish among name instances that actually belong to different authors. In contrast, FINI shows the lowest precision because the heuristic cannot distinguish instances that share the same full surname and first forename initial but really represent different authors (homonyms). Because homonyms appear in the same block, FINI always regards them to refer to the same authors. This heuristic works well for recall but degrades precision. AINI's performance in terms of precision is better than FINI because it can use more name string information (i.e., all forename initials) to distinguish name instances of different authors who do not share initialized names. However, AINI is unable to correctly distinguish among homonymous instances that share all forename initials plus the full surname but belong to different authors. The close-to-perfect precision by Author-ity2009 shows that its intricate method can successfully distinguish these challenging homonym cases by utilizing their patterns of (dis)similarity over features (e.g., coauthor names, title words, etc.). Thanks to the extremely high precision and comparatively decent recall, Autor-ity2009's disambiguation performance is stronger than baseline performances when recall and precision are weighed equally (F1).

The performance of Author-ity2009 evaluated on AUT-NIH exhibits the same patterns as those reported for AUT-ORC. One difference is that Author-ity2009 achieves almost perfect precision, recall, and F1 (> 0.99) in AUT-NIH. Another difference, however, is that the performance gaps between Author-ity2009 and baselines (Figure 6b) are smaller for AUT-NIH than they were for AUT-ORC (Figure 6a). Taken together, these two observations imply that name instances in AUT-NIH tend to be less ambiguous than those in AUT-ORC: their author identities can be matched based on initialized forenames more frequently than those in AUT-ORC (better performing baselines), while Author-ity2009 can also produce better disambiguation results for them than it does for AUT-ORC name instances. This indicates that depending on the ambiguity levels of labeled data and performance of baselines, the same performance of a disambiguation method can result in evaluation results providing slightly different impressions.

To better characterize the Author-ity2009's disambiguation performance, we turn to an analysis of its effectiveness relative to baseline methods for different name-ethnicity groups. This idea is based on the observation that certain ethnic names are more difficult to disambiguate than others (Louppe et al., 2016; Torvik & Smalheiser, 2009; Treeratpituk & Giles, 2012). Figure 7 reports $B^3$ recall and precision across ethnicity groups when Author-ity2009's disambiguation is evaluated on AUT-ORC. For visual simplicity, F1 scores are not shown.

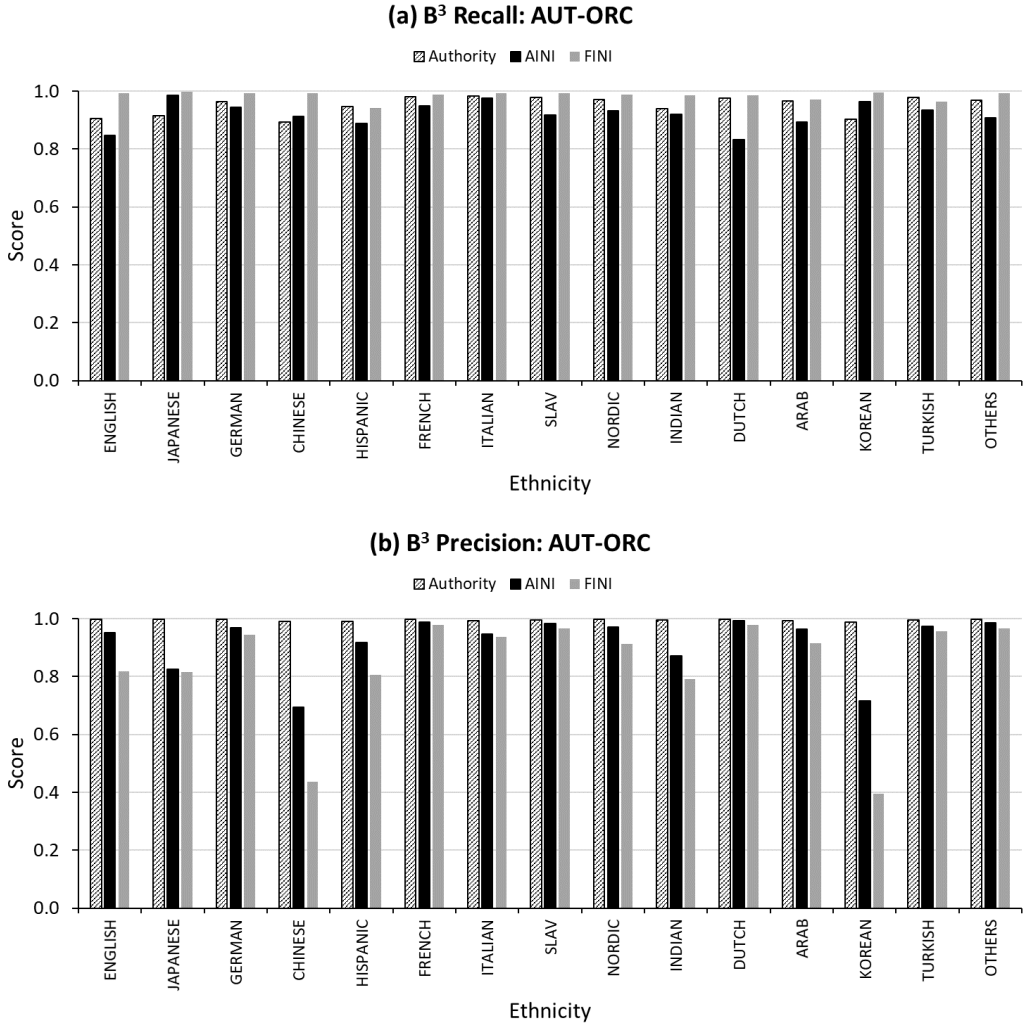

*Figure 7: Evaluation of Per-Ethnicity Disambiguation Performances of Author-ity2009 Evaluated on AUT-ORC*

Figure 7a shows that Author-ity2009 performs well for certain ethnic name instances (e.g., French, Italian, Slav, Dutch, etc.), and worse for others (e.g., English, Japanese, Chinese, Korean, etc.). FINI produces higher recall than Author-ity2009 and AINI for some ethnic names. But this comparison reveals its weak points. AINI fails to correctly decide identities of author name instances with Hispanic, Arab, and Turkish origins. These ethnic names are known to be vulnerable to becoming synonyms due to complex surnames (whose parts are often wrongly regarded as forenames) or improperly encoded characters like those from the Cyrillic alphabet (Gomide, Kling, & Figueiredo, 2017; Müller et al., 2017). On these difficult ethnic name instances, Author-ity2009 shows recall scores similar to or better than those by FINI.

The usefulness of per-ethnicity evaluation is highlighted in Figure 7b, which reports precision measures. According to Figure 7b, Author-ity2009 performs very well (> 0.99) consistently across all ethnic types. Its performance improvements over the two baseline method are most pronounced for Japanese, Chinese, Hispanic, Indian, and Korean names. For example, Chinese and Korean name instances are known to be very challenging to disambiguate, but they are correctly parsed by Author-ity2009, whereas initialized-forename-based matching strategies (AINI and FINI) fail to distinguish them in many cases.

Figure 8 shows the per-ethnicity performance of Author-ity2009 evaluated on AUT-NIH. As in Figure 6b, Author-ity2009's recall scores here are quite high, comparable to those by AINI and FINI: in Figure 8a, the heights of diagonal-lined (Author-ity2009) and gray (FINI) bars are very similar, both producing almost perfect recall scores across many ethnicities. Meanwhile, AINI (black bar) performs slightly worse than the two methods because it cannot correctly disambiguate some synonym cases. Regarding precision, however, Author-ity2009 is shown to perform very well (>0.99) across ethnicities. Especially, its performance is particularly outstanding relative to FINI and AINI for English, Chinese, Indian, and Korean names. Compared to AUT-ORC, the disambiguation performance by Author-ity2009 and baseline methods evaluated on AUT-NIH again reveal that the mix of name instances found in AUT-NIH pose fewer disambiguation challenges than those that appear in AUT-ORC. For the former, both algorithmic and heuristic methods produce almost perfect or very high (> 0.97) precision scores in distinguishing name instances of 10 (out of 15) ethnicity types. In contrast, only three ethnic groups of names (e.g., French, Slav, and Dutch) are disambiguated with similarly high precision in Figure 7b. In addition, Chinese and Korean name instances are less difficult to disambiguate using the heuristics than those in AUT-ORC: the heights of black (AINI) and gray (FINI) bars are taller than those in AUT-ORC.

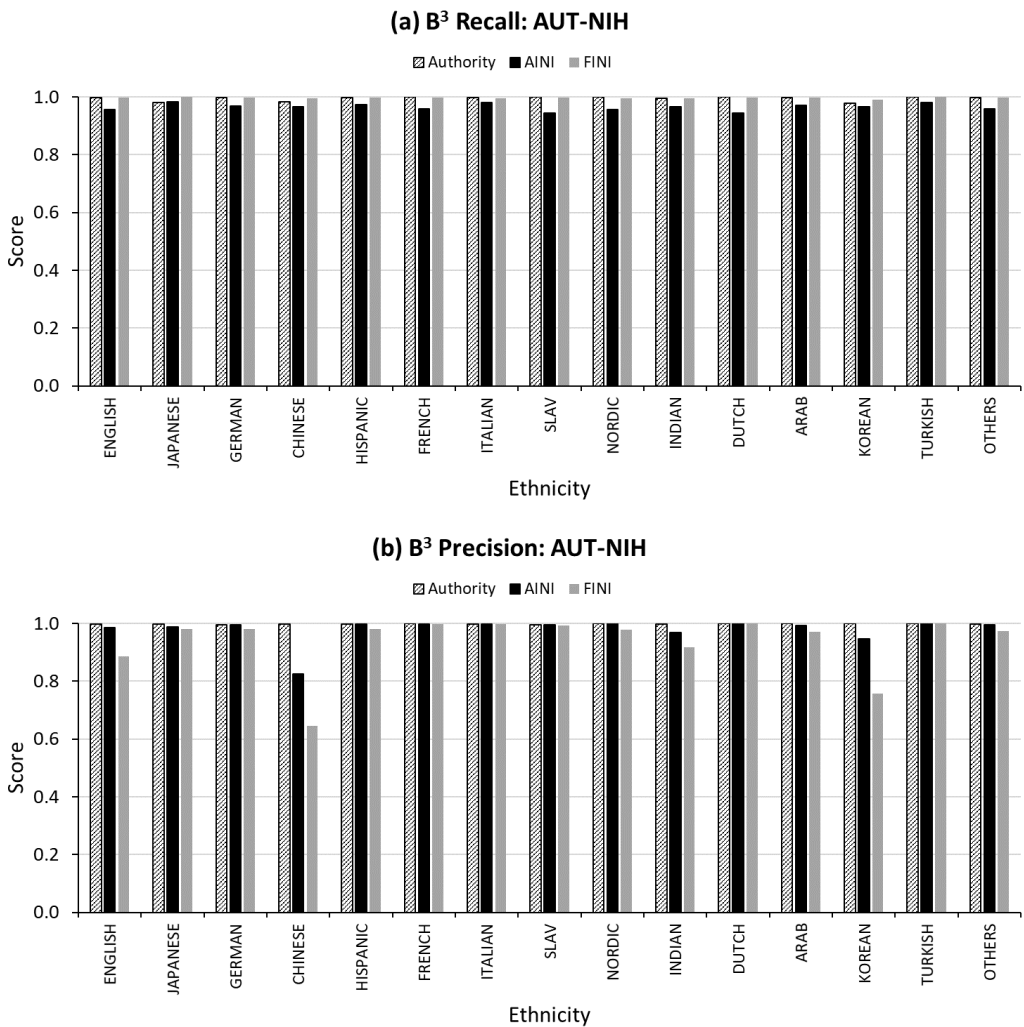

Figure 8: Evaluation of Per-Ethnicity Disambiguation Performances of Author-ity2009 Evaluated on AUT-NIH

Classification Performance: AUT-SCT

AUT-SCT consists of self-citing name instance pairs that are assumed to refer to the same authors. Because only positive matching instances can identified in these data, labeled pairs can be only used for evaluating how many pairs are correctly classified Author-ity2009 (≈ recall). Table 4 reports the accuracy of Author-ity2009 and baseline methods in classifying instance pairs in AUT-SCT.

Table 4: Accuracy of Classification Performances by Author-ity2009 and Baseline Methods Evaluated on AUT-SCT

| Disambiguation Method | Accuracy in Percentage |
|---|---|
| Author-ity2009 | 98.06 % |
| AINI | 93.59 % |
| FINI | 100.00 % |

Author-ity2009 records high accuracy (98.06%): it correctly decides a pair to represent the same author in most cases. But the simple heuristic of FINI produces a perfect accuracy score (100%). This is expected because in AUT-SCT, name instances are paired if they appear not only in citing and cited papers but also match on the full surname and first forename initial. Meanwhile, AINI produces the lowest score because it classifies a pair of instances as non-matching if they refer to the same author but have different forename initials. Following the cases of AUT-ORC and AUT-NIH, the classification performance of Author-ity2009 and baseline methods are calculated per ethnicity. The results are reported in Figure 9.

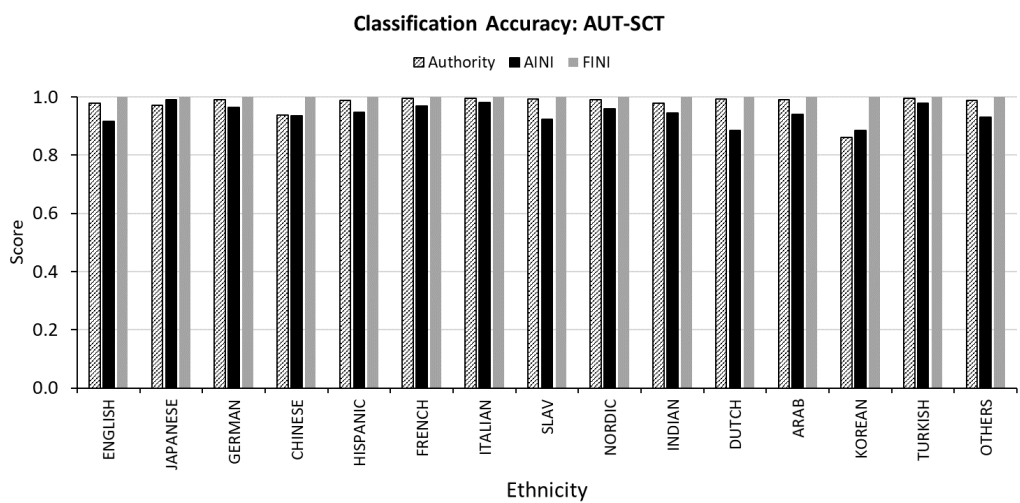

Figure 9: Evaluation of Per-Ethnicity Classification Accuracy of Author-ity2009 Evaluated on AUT-SCT

According to Figure 9, Author-ity2009 performs well in general, especially on many European ethnic names (German, Hispanic, French, Italian, etc.), but does not work as effectively on a few Asian names (e.g., Chinese and Korean). This confirms the observation in Figure 7a which shows Author-ity2009 is relatively weak in finding all name instances of distinct authors with Chinese and Korean names. Performances by AINI (black bar) show that several ethnic names such as Chinese, Dutch, English, Korean, Hispanic, and Slav are more susceptible to synonyms than other ethnic names, which confirms the observations in Figure 7a where AINI's recall performance deteriorates substantially for these ethnic names. Again, FINI (gray bar) reaches perfect scores across ethnicities as all instance pairs in self-citation relation share the same surname and first forename initial.

AUT-ORC vs. AUT-NIH vs. AUT-SCT

As reported above, three labeled datasets together highlight different aspects of Author-ity2009's disambiguation performance. They all showed that Author-ity2009 is highly accurate in disambiguating author name instances. It demonstrated special strength in distinguishing author name instances that belong to different authors and in producing almost perfect clustering precision (AUT-ORC and AUT-NIH). In addition, Author-ity2009 performed well in finding name instances of unique authors, producing very high clustering recall (> 0.96; AUT-ORC and AUT-NIH) and classification accuracy (= 98.06%; AUT-SCT) scores. Note that the Author-ty2009 is by design aimed to disambiguate with high precision because incorrectly matched name instances (merged author identities created by false positives) are more harmful than wrongly mismatched ones (split author identities created by false negatives) for bibliometric analyses (Fegley & Torvik, 2013; Liu et al., 2014; Torvik & Smalheiser, 2009). The evaluation results described so far strongly suggest that Author-ity2009 achieved its stated precision-over-recall goals. Using the name instances stratified into different ethnic groups, the three labeled datasets discussed here provide a deeper understanding of Author-ity2009's disambiguation performance. Author-ity2009 achieved high precision regardless of ethnic name types (AUT-ORC and AUT-NIH). But its recall was relatively weak in disambiguating some ethnic names, when compared with baseline performances (AUT-ORC, AUT-NIH, and AUT-SCT), suggesting possibilities to improve the algorithm.

Although three labeled datasets produced similar evaluation results, they had different characteristics. First, AUT-ORC and AUT-NIH were used to evaluate both the precision and the recall of Author-ity2009's clustering of name instances that refer to the same unique authors. But AUT-SCT could be used only to evaluate how well Author-ity2009 decided that self-citing name instance pairs refer to the same authors (≈ recall). This means that AUT-SCT could only provide partial evaluation of Author-ity2009's disambiguation performance.

Other differences between labeled datasets also help illuminate particular strengths and weaknesses in Author-ity2019. AUT-ORC and AUT-NIH have different levels of name ambiguity. When the Author-ity2009's performance was compared with two commonly-used baseline methods, it was less impressive on AUT-NIH where simpler the baseline methods accomplished equivalently high precision and recall to the more sophisticated Author-ity2009. In contrast, in AUT-ORC, the performance gaps between Author-ity2009 and baseline methods widened substantially. Considering that the baseline methods are deterministic (matching name instances on full surname and initialized forename), their strong performances mean that (1) while many name instances in AUT-ORC and AUT-NIH are not ambiguous, (2) AUT-ORC contains more ambiguous names than AUT-NIH. We observed the same patterns in comparisons of performance across groups of ethnic name instances known to vary in their ambiguity (Figure 7 and Figure 8).

Different levels of name ambiguity might arise from the different sizes of labeled data in our study: AUT-ORC contains more than 3 million instances, while AUT-NIH consists of 312K instances. As name ambiguity in bibliographic data tends to increase with data size (Fegley & Torvik, 2013; J. Kim, 2017), AUT-ORC might be naturally more ambiguous than AUT-NIH. Other differences between these datasets may result from the data sources from which they were drawn. AUT-NIH relied on funded PI information. So, the name instances that could be labelled were restricted to those of researchers who have ever received funds from NIH, a group likely to be more prominent and more homogenous than science itself. In contrast, AUT-ORC utilized ORCID profile data for more than 5 million researchers worldwide. In AUT-ORC, researcher's geo-locations were unevenly distributed (e.g., researchers in Italy and Spain are over-represented) but such an imbalance was more pronounced in AUT-NIH in which almost 55% of name instances were English origins. AUT-SCT exceeded other two labeled data by extracting more than 4 million instance pairs in self-citation relation, although its use for disambiguation evaluation is confined to measuring recall.

## Conclusion and Discussion

This study showed that the ORCID-linked labeled data can be used to evaluate the performance of a disambiguation method implemented on large-scale bibliographic data. As a use case, this study evaluated the disambiguation performance of Author-ity2009 using 3 million name instances linked to ORCID researcher profiles (AUT-ORC). For comparison, two other popular data sources - NIH-funded PI information (AUT-NIH) and self-citation information (AUT-SCT) – were also used to label name instances in Author-ity2009. Results showed that ORCID-linked labeled data better represented the gender, ethnicity and block size distributions of Author-ity2009 t (AUT-ORC > AUT-SCT > AUT-NIH), but did worse in terms of publication-year distributions (AUT-SCT > AUT-NIH > AUT-ORC), suggesting that ORCID, which skews toward younger researchers, may be more effectively used for recent disambiguation tasks.

In evaluating the clustering results of Author-ity2009, ORCID-linked labeled data effectively captured the 'high precision over high recall' strategy of Author-ity2009. Although comparative labeled data also produced the same evaluation results, ORCID-linked labeled data could provide more nuanced details about the Author-ity2009's performance when name instances were evaluated across ethnic name groups. As such, ORCID-linkage can be used as a labeling method to produce large-scale truth data to evaluate the performance of a disambiguation method from various aspects. Three large-scale labeled data – AUT-ORC, AUT-NIH, and AUT-SCT – used in this study are publicly available[17]. The data sharing is expected to assist researchers to develop, compare, and validate disambiguation models using diverse, large-scale labeled data.

This study suggests several implications for researchers and practitioners of author name disambiguation. First, ORCID can be an effective source of authority for creating labeled data. This study illustrated that ORCID-linkage can generate millions of labeled name instances in a bibliographic data, which is not easily achievable by manual or other record-linkage-based labeling. In addition, ORCID-linkage can be repeated without much additional cost once technical procedures for record-linkage are implemented. Moreover, ORCID data continue to be expanded, publicly available, and released annually. This means labeled data via ORCID-linkage can be improved in representing the population of a whole disambiguated dataset and updated on a regular basis, enabling sustained evaluation of author name disambiguation in ever-growing digital libraries.

Second, ORCID-linked labeled data can complement other types of linkage-based labeled data. Our comparisons across three different types of linked labeled data showed that , ORCID-linked labeled data could captured the aspects of Author-ity2009's performance that were also identified in the other two datasets. This means ORCID linkage can be used as an alternative to other labeling methods if they are unavailable. In addition, ORCID-linkage can be used to help researchers evaluate the labeling quality of other labeled data. Out of 312,951 instances in AUT-NIH, for example, a total of 32,131 instances were also linked to 3,578 ORCID ids. Among them, 99 name instances were assigned to different authors by the ORCID-linkage and the NIH-ExPORTER linkage used in Lerchenmueller and Sorenson (2016).

---

[17] Datasets can be downloaded at https://doi.org/10.6084/m9.figshare.13404986.v1

Manual inspection using online researcher profiles revealed that 90 instances were wrongly assigned by the NIH-ExPORTER linkage, while 9 instances were mistaken by the ORCID-linkage.

Third, ORCID-linked labeled data can provide more enriched evaluation results. They can be used together with other labeled data for triangulating a disambiguation method's performance. Unlike self-citation-based labeled data, ORCID-linked labeled data can be used to measure both clustering and classification performances. Unlike NIH-linked labeled data, ORCID-linked labeled data contain a greater range of ambiguous names across ethnicities, which can enable a disambiguation method to be evaluated on name instances with different ambiguity levels. This in turn allows for more focused analysis to address difficult disambiguation tasks such as those presented by synonyms and homonyms. Moreover, ORCID-linkage can produce labeled instances that are challenging to disambiguate but are not easily collectable by other labeling methods. For example, FINI could not reach perfect recall in AUT-ORC (Figure 6a and Figure 7a). As detailed above (see Clustering Performance: AUT-ORC and AUT-NIH), 273,782 name instances of 12,646 authors (= unique ORCID ids) are recorded in a way that their 'surname + first forename initial' strings of the same author are different. This means ORCID-linkage could produce labeled name instances that refer to the same authors but do not belong to the same blocks. Such synonymous name variants existing across blocks have been insufficiently studied in disambiguation research (Backes, 2018; Gomide et al., 2017) because many studies have created labeled data by collecting (= blocking) ambiguous name instances sharing at least the full surname and first forename initial (J. Kim, 2018; Müller et al., 2017). Using ORCID-linked labeled data, scholars can develop disambiguation models that address synonyms as well as homonyms.

Furthermore, ORCID-linkage can help researchers label the name instances of authors who work in diverse research fields for which labeled data are scarce. Most existing labeled datasets for author name disambiguation were created to disambiguate author names in a few scientific domains, especially Computer Science and Biomedical Sciences (A. A. Ferreira, Gonçalves, & Laender, 2012; Müller et al., 2017). For those who need to mine ambiguous bibliographic data that represent diverse fields, ORCID-linkage can be an effective way to generate labeled data for their ad-hoc disambiguation tasks.

To promote the use of ORCID as a labeling source for author name disambiguation, however, several issues need to be addressed. First, our discussion of representativeness shows that name instances labeled through ORCID linkage may not generalize to the population of scientists because it over-represents early and mid-career researchers and underrepresents Asian names. This implies that ORCID cannot eliminate the need for author name disambiguation in bibliographic data until it becomes a universal author identification system. The same issue occurred to other labeled data in this study. To mitigate the problem, stratified sampling of name instances may be considered to create a set of labeled name instances that represent better population data.

Second, the accuracy of ORCID records still needs to be verified. As acknowledged by ORCID, some records may contain errors due to "benign" (unintentional) mistakes by profile creators (e.g., claiming other researcher's work as their own)[18]. Note that other labeled data may have the same verification problems. Human experts can produce inaccurate labels and often disagree on labeling decisions even given the same information (Shin et al., 2014; Song et al., 2015). Although NIH PI data are curated with special care by NIH, the linkage process for labeling may entail erroneous matching between PI names and author names in NIH-funded papers. As shown above regarding the labeling quality of AUT-NIH (see 3rd paragraph in Conclusion and Discussion), ORCID-linked data provided more accurate labeling results than the other method but still contained erroneous labels. To ensure that errors in ORCID records do not affect disambiguation evaluation, the accuracy of ORCID records may be tested on various samples or sensitivity analyses may be conducted to find how many errors in ORCID-linked labeled data are acceptable for robust evaluation results[19].

---

[18] https://qa.orcid.org/node/68

Third, the completeness of ORCID's coverage of a researcher's publications needs to be investigated. As pointed out in Youtie et al. (2017), the ORCID publication list of a researcher may be incomplete due to, for example, lack of timely updates. Finding any systematic patterns of incomplete coverage may enable us to better understand the characteristics of ORCID to enhance its usefulness as a labeling source for evaluating author name disambiguation at scale.

Appendix A

One concern regarding the evaluation procedure in this study is that the ethnicity and gender tagging may be inaccurate. The tools – *Ethnea* and *Genni* – used for ethnicity and gender predictions in this study are reported to produce more accurate and less missing prediction results than other existing tools at the time of their publication (Mishra, Fegley, Diesner, & Torvik, 2018; Torvik & Agarwal, 2016). However, several tools have recently showed that they outperform previous techniques including *Ethnea* and *Genni* (e.g., Santamaría & Mihaljević, 2018; Ye et al., 2017). However, we believe *Ethnea* and *Genni* are adequate tools for gender and ethnicity predictions to group name instances in MEDLINE for evaluating author name disambiguation by Author-ity2009 because their prediction models were built and validated based on the MEDLINE data. For example, the high-performing tool for ethnicity prediction in Ye et al. (2017) shows very promising prediction results on Wikipedia and Email/Twitter data but it is unknown how it would perform on author name instances in MEDLINE which are the target of ethnicity prediction for this study.

We conducted a sensitivity analysis on the distribution of ethnicity to see how it would change if ethnicity prediction results are different. For this, we randomly changed ethnicity tags of name instances that constitute 10% of each ethnic groups and re-run the evaluation procedures. The 10% random selection is based on the performance differences between *Ethnea* and *NamePrism* reported in Ye et al., (2017). The results show that the ethnicity tagging errors indeed changed the distributions (5~12% differences in ratios depending on ethnicities). However, performance evaluation results for Author-ity2009 in comparison with baselines (AINI and FINI) over different linked data (AUT-ORC, AUT-NIH, and AUT-SCT) did not change much. Interestingly, induced tagging errors, the performance gaps between Author-ity2009 and baselines were shown to get widened for many ethnicities. We conjecture that highly ambiguous names such as Chinese, English, and Korean were wrongly tagged as other less ambiguous ethnicities, which increased the name ambiguity level of the ethnicity groups and decreased the performances of baseline disambiguation on them.

Acknowledgement

This paper was supported by funds from the National Science Foundation (#1917663, #1760609, and #1561687). We would like to thank anonymous reviewers whose comments and suggestions helped us improve this paper.

---

[19] Self-citation pairs are prone to errors due to homonym pairs (same name strings but different authors).